\documentclass[prl,aps,epsfig,superscriptaddress,twocolumn]{revtex4}
\usepackage{bm}
\usepackage{times}
\usepackage[dvips]{graphicx}
\usepackage{mathrsfs}
\usepackage[intlimits]{amsmath}
\usepackage[colorlinks, citecolor=red]{hyperref}

\begin{document}

\title{Experimental demonstration of memory-enhanced scaling for entanglement connection \\ of quantum repeater segments}

\author{Yun-Fei Pu$^1$\footnote[2]{These authors contributed equally to this work.}\footnote[3]{Present address: Institute for Experimental Physics, University of Innsbruck, A-6020 Innsbruck, Austria}, Sheng Zhang$^{1}$\footnotemark[2], Yu-Kai Wu$^1$, Nan Jiang$^1$\footnote[4]{Present address: Department of Physics, Beijing Normal University, Beijing 100875, China}, Wei Chang$^1$, Chang Li$^1$, Lu-Ming Duan\footnote[1]{Corresponding author: lmduan@tsinghua.edu.cn}}

\affiliation{Center for Quantum Information, IIIS, Tsinghua University, Beijing 100084, PR China}

\begin{abstract}
The quantum repeater protocol is a promising approach to implement long-distance quantum communication and large-scale quantum networks. A key idea of the quantum repeater protocol is to use long-lived quantum memories to achieve efficient entanglement connection between different repeater segments with a polynomial scaling. Here we report an experiment which realizes efficient connection of two quantum repeater segments via on-demand entanglement swapping by the use of two atomic quantum memories with storage time of tens of milliseconds. With the memory enhancement, scaling-changing acceleration is demonstrated in the rate for a successful entanglement connection. The experimental realization of entanglement connection of two quantum repeater segments with an efficient memory-enhanced scaling demonstrates a key advantage of the quantum repeater protocol, which makes a cornerstone towards future large-scale quantum networks.
\end{abstract}

\maketitle

\section{Introduction}

Long-distance quantum communication and large-scale quantum networks constitute one of the central tasks in quantum information science\textsuperscript{\citenum{internet}}. Direct communication of quantum signals in optical fibers are hindered by the inevitable exponential loss of photons with the communication distance. The quantum repeater protocol provides a promising approach to solve this problem\textsuperscript{\citenum{BDCZ,DLCZ,Gisin}}, where long-distance communication is established through entanglement connection of many smaller segments of communication channels and the exponential growth of noise is suppressed through heralding and nested entanglement purification\textsuperscript{\citenum{BDCZ,teleportation,purification}}. A well-known approach to the implementation of quantum repeaters is the Duan-Lukin-Cirac-Zoller (DLCZ) scheme\textsuperscript{\citenum{DLCZ}}, where collective spin excitations in atomic ensembles\textsuperscript{\citenum{Polzik}} are employed to provide the required quantum memory and the heralded entanglement connection is used to boost the scaling of efficiency through the memory enhancement. Many impressive experimental advances have been reported along this approach\textsuperscript{\citenum{7,8,9,10,Kimble2007,11,pan2008,pan50km}}. Entanglement generation between two quantum memories (repeater nodes) has been reported using atomic or spin ensembles\textsuperscript{\citenum{Kimble2007,11,pan2008,pan50km}} as well as single atomic ions or diamond defect spins\textsuperscript{\citenum{Monroe, Lukas, Rainer, Weinfurter,Hansonphoton}}. Memory enhancement in efficiencies has been observed in heralded entanglement generation\textsuperscript{\citenum{Kimble2007,Kwiat}} by employing atomic or optical quantum memories. Memory-accelerated quantum key distribution has been demonstrated recently\textsuperscript{\citenum{lukinnature}} by sending weak coherent pulses to a middle node for the Bell-state measurement. There is also an alternate approach towards realization of quantum repeaters with an all-photonic scheme where the requirement of the quantum memory is replaced by preparation of a large repeater graph state\textsuperscript{\citenum{photonic_repeater}} with noticeable experimental progress recently\textsuperscript{\citenum{pan19repeater,Hasegawa19repeater}}. For the memory based approach to quantum repeaters, a goal that remains outstanding is to demonstrate scaling change in efficiencies enabled by quantum memories for entanglement connection of two quantum repeater segments, which is the key ingredient for the quantum repeater protocol to achieve its efficient scaling.

In this paper, we report an experimental realization of entanglement connection between two quantum repeater segments with a quantum-memory-enhanced scaling for its connection efficiencies. Compared with the direct entanglement swapping of two synchronous entangled photon pairs without the use of quantum memories\textsuperscript{\citenum{PanPRL2003}}, we demonstrate that the efficiency in entanglement connection of two segments improves from a quadratic scaling to a linear scaling with the preparation efficiency of each entangled pair. This change in scaling of efficiencies, when extended to multiple segments, is the key for the quantum repeater protocol to improve from an exponential scaling to a polynomial scaling\textsuperscript{\citenum{BDCZ,DLCZ,Gisin}}. A challenging requirement for the demonstration of the scaling change in entanglement connection is that the coherence time of the quantum memories has to be longer than the preparation and the successful heralding time of each entangled pair. In our experiment, with the help of optical traps, we have two atomic quantum memories with the storage time of tens of milliseconds, long enough for the on-demand entanglement swapping. Atom-photon entanglement is generated asynchronously in two long-lived quantum memories, and the entanglement swapping between them is implemented on demand only when both sides have successfully registered a photon (in general at different times). This experiment, as the first demonstration of the scaling change in efficiencies for entanglement connection of two quantum repeater segments, provides a key enabling ingredient for the realization of larger-scale quantum repeaters and quantum networks.

\begin{figure*}
  \centering
  \includegraphics[width=16cm]{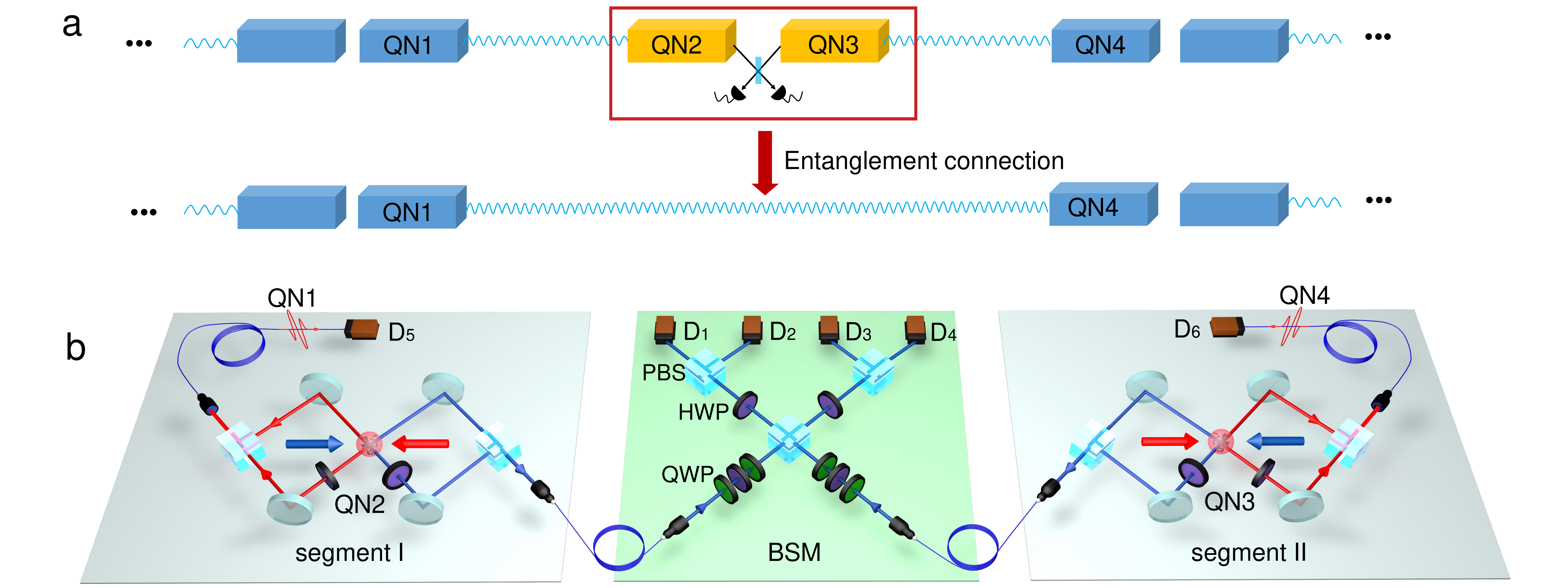}\\
  \caption{ The quantum repeater protocol and the experimental setup.
   a, A sketch of entanglement connection (swapping) in the quantum repeater protocol. QN represents quantum repeater nodes. b, The whole experimental setup consists of three parts: segment I (QN1 and QN2) and segment II (QN3 and QN4), together with a Bell state measurement (BSM) station in the center. QN2 and QN3 are two similar atomic memory nodes separated by $3\,$m in space. QN1 and QN4 are photons in this experiment and are measured by detectors D5 and D6. A sandwich structure consisting of a quarter-wave plate (QWP), a half-wave plate (HWP) and another QWP is introduced to compensate the polarization change in the fiber transmission. The coincidence events between the single photon detectors D1 and D4 (or D2 and D3) project the two idler photons onto one of the four Bell states $|\Phi^{+}\rangle$.
  }
\end{figure*}

As illustrated in Fig.~1a, the key idea of the quantum repeater architecture is to divide the communication channel into multiple segments, to establish entanglement for each segment which is stored in quantum memories, and then to use entanglement swapping to connect these segments when each segment of entanglement has been successfully heralded. Here we consider the demonstration of this architecture for the primitive case with two segments. As the two end nodes QN1 and QN4 do not need further connection, they can be measured according to the application protocol (such as for the entanglement-based quantum key distribution\textsuperscript{\citenum{E91,DLCZ}}) before the connection of two middle nodes QN2 and QN3. Therefore, in this simplified configuration, we do not need a quantum memory for the end nodes QN1 and QN4. In our experiment, we implement the qubits at QN1 and QN4 with photons and the qubits at QN2 and QN3 with atomic quantum memories as shown in Fig.~1b. We generate atom-photon entanglement for each segment and perform on-demand entanglement swapping on the atomic memory qubits QN2 and QN3 through the Bell measurement only when the detectors D5 and D6 at the nodes QN1 and QN4 have both successfully heralded registration of a photon count. Note that different from the configuration in Ref.\textsuperscript{\citenum{pan2008}} which generates entanglement between two edge atomic memory nodes by simultaneous heralding of two photon counts in the middle, it is critical here that the middle nodes QN2 and QN3 need to be quantum memories so that we can store heralded states prepared asynchronously in these memories for an efficient entanglement connection.

\section{Results}

In both QN2 and QN3, coherence time of tens of milliseconds inside $^{87}$Rb atomic ensembles is achieved through spatial confinement with an optical lattice\textsuperscript{\citenum{zhaoran08,yangsjlong,magic}}. A one-dimensional (1D) optical lattice holds the atoms and suppresses their motional decoherence by confining the atoms inside a single pancake potential of the lattice. The clock states ($|g\rangle \equiv |5S_{1/2},F=2, m_F=0\rangle \leftrightarrow |s\rangle \equiv |5S_{1/2},F=1, m_F=0\rangle$) are used to minimize the influence of the temporal and spatial fluctuation noise of the magnetic field\textsuperscript{\citenum{zhaoran08}}, and a magic-value magnetic field of $B\approx 4.3\,$G is employed to cancel the differential AC-Stark shift induced by the lattice intensity inhomogeneity\textsuperscript{\citenum{magic,yangsjlong}}, as shown in Fig.~2a. The short-term ($\le 1\,$ms) storage efficiency of QN2 is measured by the DLCZ scheme, as shown in Fig.~2b. The fast decay in this period is induced by the atomic motion inside a single lattice pancake\textsuperscript{\citenum{zhaoran08,yangsjlong}}. The long-term ($>1\,$ms) lifetime of the quantum memory in each node is measured by the electromagnetically induced transparency (EIT) storage\textsuperscript{\citenum{magic,EIT}}, which is described in detail in the Supplementary Material. As illustrated in Fig.~2c, the long-term EIT-storage time is $77(3)\,$ms and $14(1)\,$ms for QN2 and QN3, respectively. We attribute the large discrepancy between the lifetime of the two memories to the imperfect geometry of the probe, the control and the lattice beams, which induces imperfect suppression of motional decoherence, as well as the instability of the magnetic field due to the power supply noise at QN3.

We use a variation of the DLCZ scheme to generate the atom-photon entanglement in each segment. The quantum information in the atomic ensemble is carried by the two magnetic-field insensitive levels $|g\rangle$ and $|s\rangle$ in the ground state manifold\textsuperscript{\citenum{zhaoran08, DLCZ}}. After the atoms are loaded into the 1D optical lattice and are optically pumped to the initial state $|g\rangle$, a weak write pulse in linear polarization drives a Raman transition from $|g\rangle$ to $|s\rangle$. We collect photons in two symmetric signal modes $S_L$ and $S_R$ with angles of $\pm 1.5^\circ$ relative to the write beam, and two spatial modes of the spin-wave excitation L and R are defined correspondingly in the atomic cloud\textsuperscript{\citenum{pan2008}}. After we combine the two signal photon modes $S_L$ and $S_R$ on a polarizing beam splitter PBS2 as shown in Fig.~2a and ignore the small higher-order excitation terms, the resulting atom-photon entangled state can be expressed as (with the vacuum part neglected as it will be eliminated by the subsequent measurements):
\begin{equation}
|\Psi\rangle _{S-A}=\frac{1}{\sqrt{2}}(|H\rangle|L\rangle+e^{i\phi_S}|V\rangle|R\rangle), \label{EQ1}
\end{equation}
where $|H\rangle$/$|V\rangle$ represents the horizontal/vertical polarization of the signal photon, $|L\rangle$/$|R\rangle$ represents a single collective excitation in the spatial mode L/R, and $\phi_S$ is the phase difference between the two signal paths before they are combined on the PBS2.

To evaluate the quality of the generated atom-photon entanglement in each segment, we coherently convert the spin-wave $|L\rangle$ and $|R\rangle$ into two idler photon modes $I_L$ and $I_R$ with a read pulse resonant to the transition $|s\rangle \rightarrow |e\rangle$ propagating in the opposite direction to the write beam. After the two idler modes are combined on the PBS1 (see Fig.~2a), the signal-idler photon state can be expressed as $|\Psi\rangle _{S-I}=\frac{1}{\sqrt{2}}(|H\rangle_S|H\rangle_I+e^{i(\phi_S+\phi_I)}|V\rangle_S|V\rangle_I),$ where $\phi_I$ is the propagation phase difference in the paths $I_L$ and $I_R$. To keep the total phase $\phi_S + \phi_I$ constant, we actively stabilize the path difference between L and R. Through quantum state tomography, the density matrix of this entangled state is reconstructed via the maximum likelihood method, and the fidelity with a two-qubit maximally entangled state is over $90\%$ for both QN2 with a storage time varying from $10\,\mu$s to $1\,$ms, and QN3 at $10\,\mu$s (see Fig.~2b). We attribute the infidelity mainly to the imperfect optical pumping and the signal-to-noise ratio in the retrieval process of the idler photon.

\begin{figure}
  \centering
  \includegraphics[width=8.7cm]{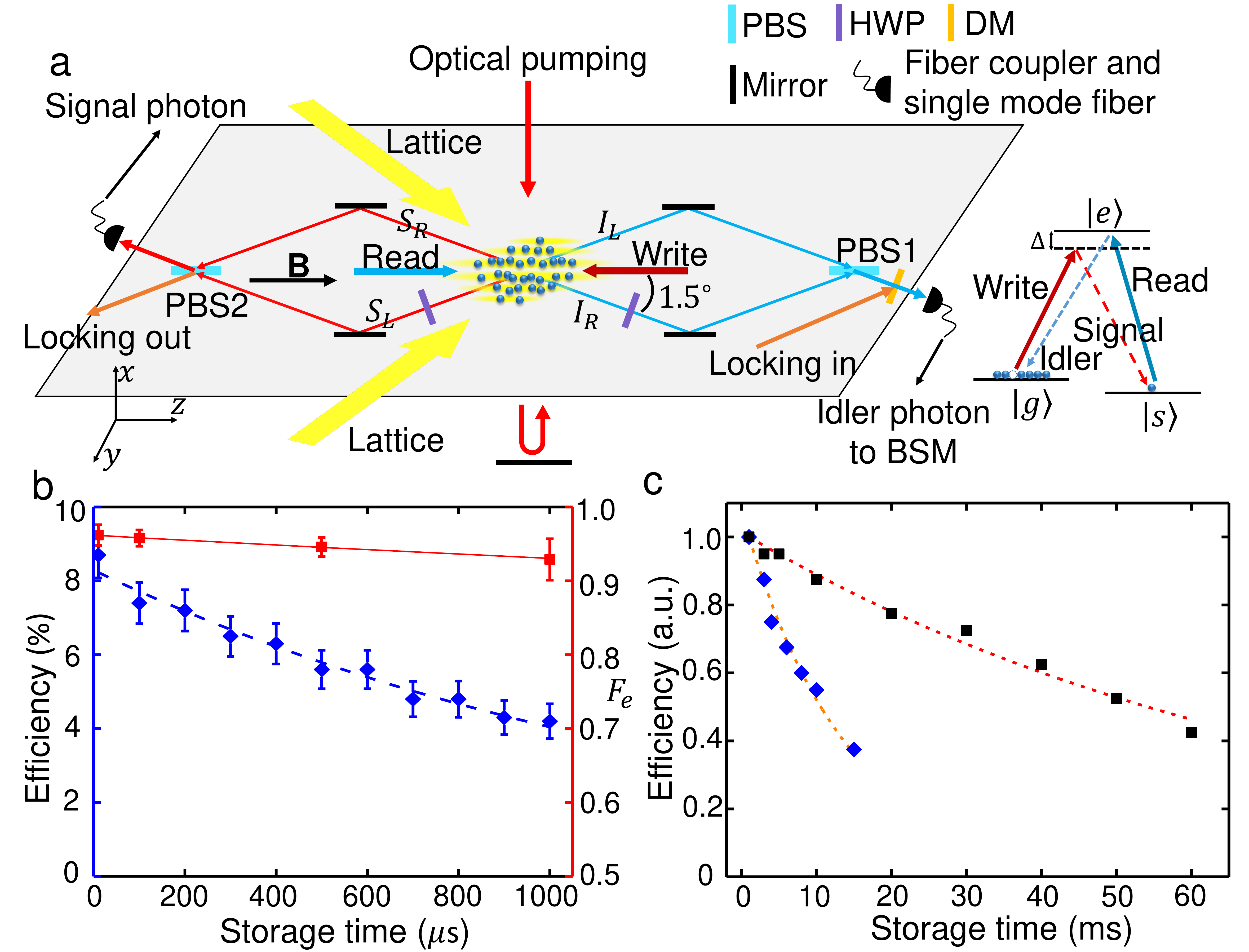}\\
  \caption{ The quantum repeater node with long memory time.
  a, A zoom-in of QN1 (a signal photon) and QN2 (an atomic quantum memory). b, The short-term overall retrieval efficiency (blue diamond) and the atom-photon entanglement fidelity $F_e$ (red square) within $1\,$ms storage in QN2. In both plots the error bars represent one standard deviation. c, The long-term storage efficiency measured by EIT storage at QN2 (black square) and QN3 (blue diamond) normalized to the time point of $t=1\,$ms. Fitted with an exponential function (red/orange dashed lines), the $1/e$ lifetime of QN2 and QN3 are $77(3)\,$ms and $14(1)\,$ms, respectively.}
\end{figure}

The protocol in this experiment can be split into three steps: the asynchronous preparation of atom-photon entanglement in segment I (QN1 and QN2) and segment II (QN3 and QN4), and the following entanglement connection, as shown in Fig.~3a. In step I, we generate atom-photon entanglement between QN1 and QN2 with a write-clean pulse train repeated at a rate of $1\,$MHz (see Supplementary Material). Once a signal photon is recorded by the detector, the write-clean cycle halts and the sequence enters step II immediately.

In step II, we generate the atom-photon entanglement between QN3 and QN4 in a similar way as in step I. Here the difference is that once the number of write-clean trials reaches $1000$, which is about $1\,$ms after the generation of atom-photon entanglement in segment I, the whole sequence will terminate and restart from step I. This $1\,$ms window is determined as a trade-off between the near-deterministic detection of a signal photon at QN4 and the low decoherence of the established atom-photon entanglement in segment I during this window. As an example, with a medium overall signal photon detection probability $p=0.3\%$ which corresponds to an intrinsic excitation probability of $\chi \approx 1\%$, $1000$ trials yield a nearly deterministic success probability of $1-(1-0.3\%)^{1000} \approx 95\%$ to generate the atom-photon entanglement. On the other hand, with the long coherence time of QN2, the atom-photon entanglement fidelity in segment I changes only slightly in the $1\,$ms window, and remains as high as $92.9(2.8)\%$ at $1\,$ms, as illustrated in Fig.~2b. Although the retrieval efficiency decays to about one half of the initial value within the first $1\,$ms, it only influences the entanglement swapping efficiency by a constant factor, which will not change the scaling. After the successful detection of a signal photon at QN4, the sequence goes to step III.

In many previous atom-photon entanglement experiments where two Zeeman sub-levels are employed as the bases of stationary qubits\textsuperscript{\citenum{yangsjzeeman, Monroe, Lukas, tracy}}, the phase in the established atom-photon entangled state oscillates due to different Larmor precession rates of these two bases. For the asynchronous connection of two segments in a quantum repeater protocol, significant problems will appear due to the random entanglement preparation time between adjacent segments and hence the random accumulated phase. Here in our experiment, the two bases of the stationary qubit $|L\rangle$ and $|R\rangle$ are nearly identical except the spatial orientation of the spin-wave vectors which have no influence on the phase\textsuperscript{\citenum{pan2008}}. Thus our established atom-photon entanglement (Eq.(\ref{EQ1})) in each node has a time-independent phase $\phi_S$ (see Supplementary Material), which guarantees that the final entangled state after entanglement connection will have a constant phase no matter when the entanglements are prepared in the two segments.

After the successful detection of two signal photons in step I and II, we connect the two segments through entanglement swapping and entangle QN1 and QN4 (two signal photons) in a post-selected way in step III. We coherently convert the spin-wave qubits in the two memories to idler photons, which are further directed to the middle station to perform the joint Bell state measurement in the photonic polarization basis (see Fig.~1b). After the successful entanglement swapping between QN2 and QN3, the two signal photons are projected into a maximally entangled state
\begin{equation}
|\Psi\rangle _{S-S}=\frac{1}{\sqrt{2}}(|H\rangle_1|H\rangle_4 + e^{i(\phi_{\text{I}}+\phi_{\text{II}})}|V\rangle_1|V\rangle_4) \label{EQ2}
\end{equation}
where $H(V)_{1(4)}$ represents a signal photon in the horizontal (vertical) polarization basis at QN1(4), and $\phi_{\text{I}}+\phi_{\text{II}}$ is a constant phase including all the path phases in the write and the read processes in both segments. We perform quantum state tomography on this signal photon entangled state by setting proper polarization measurement bases before the detectors D5 and D6, and reconstruct the density matrix with the maximum likelihood method (Fig.~3b). With the atom-photon entanglement generation probabilities in both segment I and II set to $0.1\%$, the fidelity of the reconstructed signal-signal entangled state to a maximally entangled state is measured to be $79.5(4.8)\%$, which is a clear evidence of the generated entanglement between QN1 and QN4 and confirms the success of the whole protocol.

\begin{figure}
  \centering
  \includegraphics[width=8.7cm]{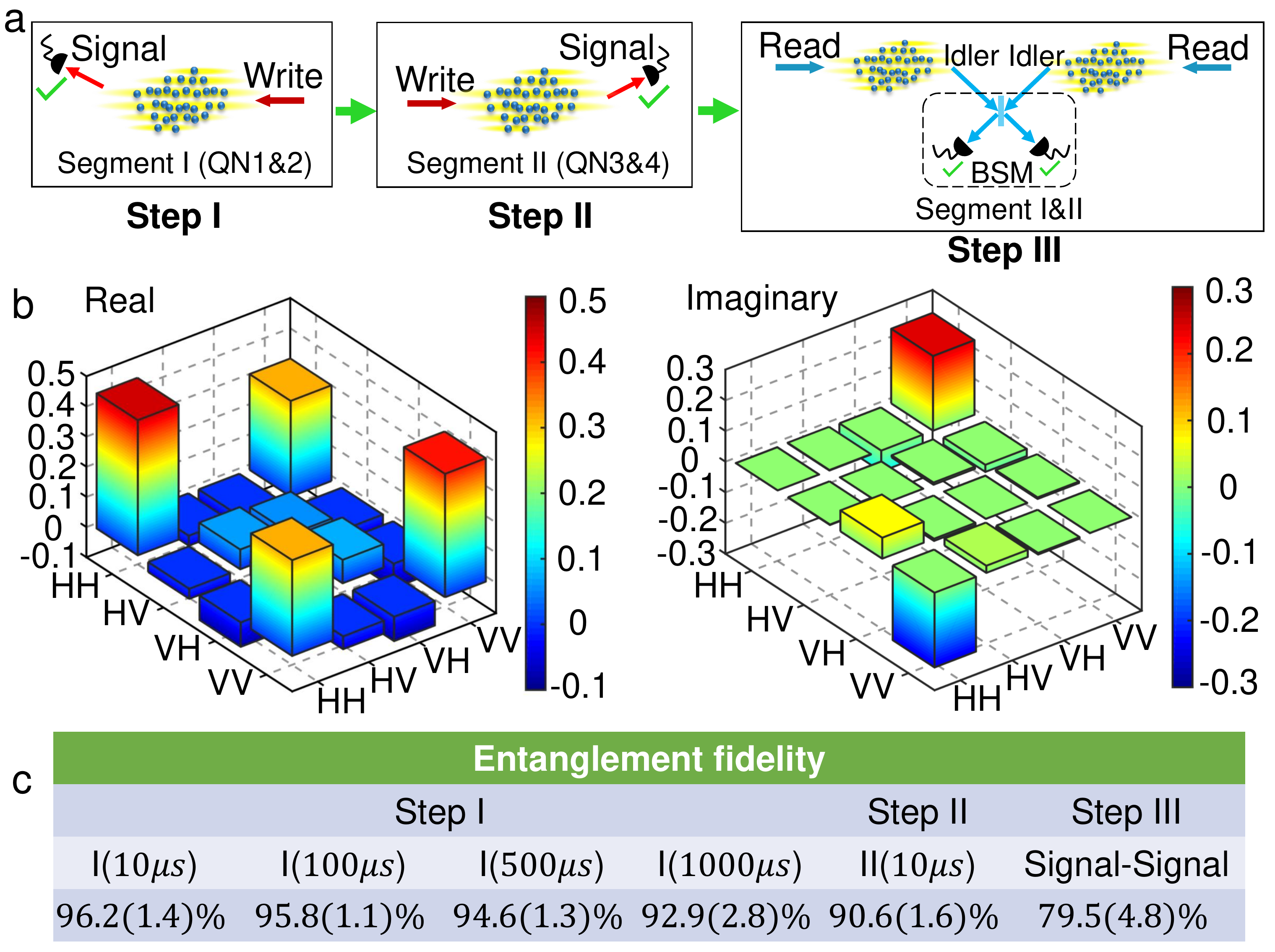}\\
  \caption{ Entanglement connection of two quantum repeater segments.
  a, The protocol of the asynchronous generation and connection of entanglement. b, The reconstructed density matrix of the signal-signal state after entanglement swapping. The fidelity of the reconstructed state to the nearest maximally entangled state is $79.5(4.8)\%$. The density matrix shown here represents the average over all the entangled states from $1000$ trials. Although the quality of the final entanglement depends on how long the heralded state is stored in QN2, this dependence is weak as the coherence loss in QN2 is small in this experiment. The whole data acquisition time is $16$ hours for all the $16$ measurement bases of the final QN1-QN4 state, and the total number of the recorded four-photon coincidence count is $656$.  c, The atom-photon entanglement fidelity in segment I after a storage of 10$\,\mu$s, 100$\,\mu$s, 500$\,\mu$s, 1000$\,\mu$s, and in segment II after a storage of 10$\,\mu$s, together with the final signal-signal entanglement fidelity. Error bars represent one standard deviation.  }
\end{figure}

\begin{figure}
  \centering
  \includegraphics[width=6cm]{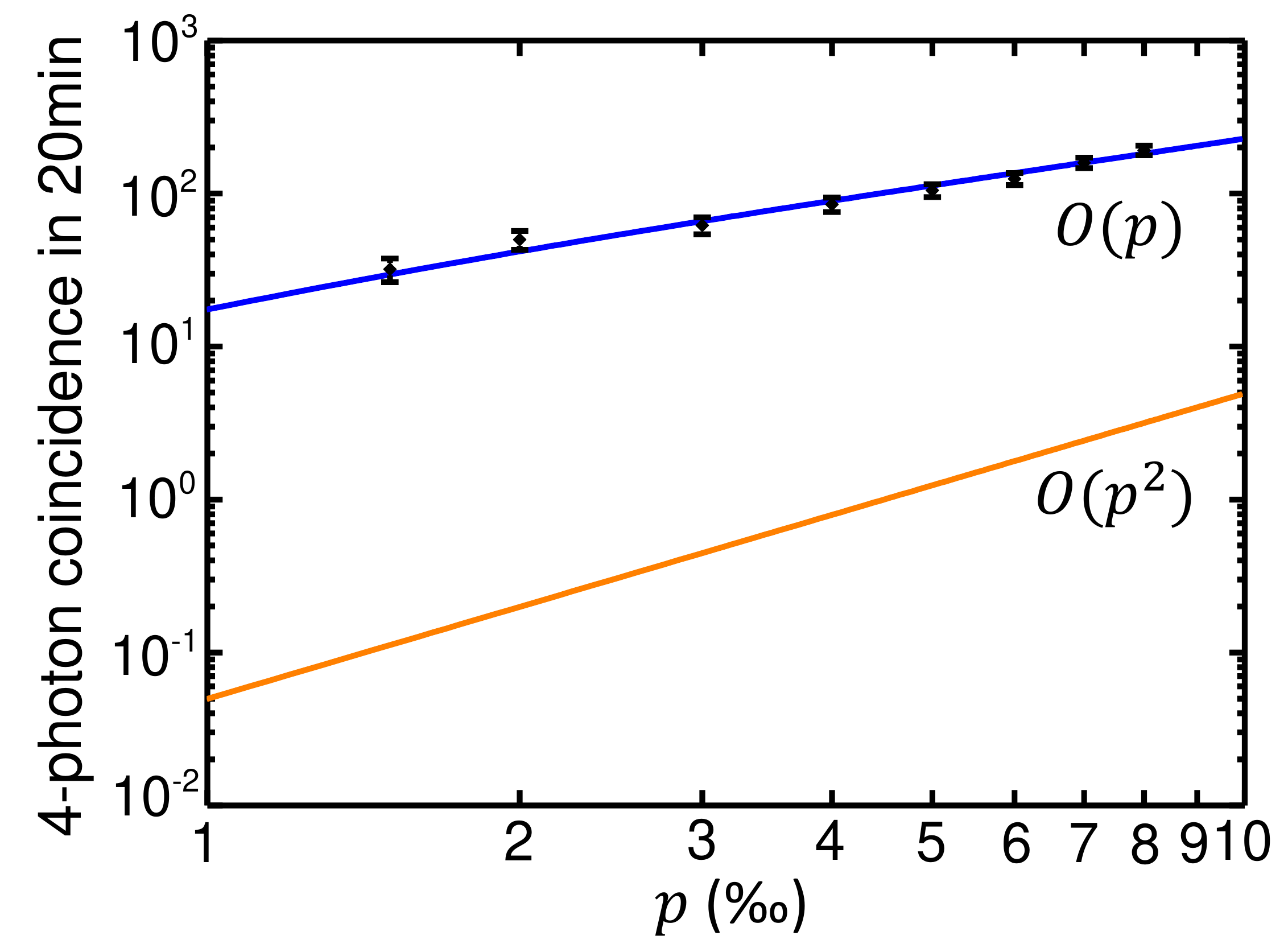}\\
  \caption{ Memory-enhanced scaling of efficiencies for entanglement connection.
   Four-photon coincidence counts in 20 minutes are measured at an excitation probability of $p$ varying from 0.15\% to 0.8\% (black diamonds). Error bars represent one standard deviation. The blue solid curve is the theoretical four-photon coincidence rate in our experiment and the orange curve is the calculated four-photon rate of the same entanglement swapping protocol but without use of the quantum memories (see Supplementary Material). For these two cases (with or without use of the quantum memories), the fidelity of the final state should be comparable as the additional fidelity decay induced by the coherence loss in the quantum memory, which is about $3\%$ for the required maximal 1 ms storage time from Fig.~3c, is much less than the overall infidelity of the final state, measured to be around $20\%$ and mainly from the contribution of the technical noise in entanglement swapping and the addition of the infidelities from the two-segment EPR states. However, the efficiency improves from a quadratic dependence ($O(p^2)$) to a linear scaling ($O(p)$) by use of the memory enhancement.}
\end{figure}

Finally we investigate the scaling property of our protocol. Suppose the success probability of generating atom-photon entanglement on both sides are set to $p$. If $p>1/n$, where $n$ is the maximally allowed trial number in the write process of segment II, the atom-photon entanglement in segment II can be prepared nearly deterministically. Thus the expected number of trials to successfully generate the entanglement on both sides and to perform a successful entanglement swapping is proportional to $2C/p$, which is linearly dependent on $2/p$ with a varying coefficient $C(p,n)\in (1, 1.42)$ (see Supplementary Material); in comparison, without the memory enhancement, the successful entanglement connection will require simultaneous entanglement generation on both sides and therefore the number of trials scales as $1/p^2$. Here we verify that the overall success rate of entanglement generation and connection between the two segments is linearly dependent on $p$ (that is, the expected time or trial number is $O(1/p)$, where $O(1/p)$ denotes the order of $1/p$) by measuring the four-photon coincidence rate with the two signal photons in $|H\rangle_{1}$ and $|H\rangle_{4}$ and the two idler photons detected in the BSM, which is one half of the signal-signal entanglement generation rate in our case (see Eq.(\ref{EQ2})). As shown in Fig.~4, when $p$ is varied, the signal-signal entanglement is generated at a rate linearly dependent on the excitation probability $p$ in each segment, which confirms the fundamental acceleration in entanglement connection from a quadratic scaling to a linear scaling for the two-segment case. In Fig.~4, we also compare the four-photon coincidence rates between this memory-enhanced protocol and a corresponding entanglement swapping protocol without use of the quantum memories (see Supplementary Material), and we can see that the advantage of the memory-enhanced protocol is more evident at a smaller $p$ due to the improved scaling with $O(p)$.

\section{Discussion}

In summary, we have demonstrated entanglement connection of two quantum repeater segments with improved scaling in efficiencies through the enhancement by quantum memories. In future, one can combine this setup with the frequency conversion setup to convert the photons on each side to the telecom wavelength so that the communication length of each segment can be extended to tens of kilometers\textsuperscript{\citenum{pan50km,Ben,Chang}}. The scaling change from a quadratic dependence on the generation rate of each segment to a linear dependence can then be used to significantly boost the overall communication efficiency, which is a key advantage of the quantum repeater protocol compared with the direct communication scheme. This work thus demonstrates an important primitive of the quantum repeater protocol and provides a building block for the implementation of long-distance quantum communication and large-scale quantum networks.

\section{Methods}

Methods are included in the Supplementary Material.

\section{Data availability}

The data that support the plots within this paper and other findings of this study are available from the corresponding author upon reasonable request

\section{Code Availability}

The code used for quantum state tomography is available from the corresponding author on reasonable request.

\section{References}

\makeatletter
\renewcommand\@biblabel[1]{#1. }
\makeatother

\section{Acknowledgements}

This work was supported by the National key Research and Development Program of China (2016YFA0301902), the Beijing Academy of Quantum Information Sciences, the Frontier Science Center for Quantum Information of the Ministry of Education of China, and the Tsinghua University Initiative Scientific Research Program. Y.K.W. acknowledges support from Shuimu Tsinghua Scholar Program and the International Postdoctoral Exchange Fellowship Program.

\section{Author Contributions}

L.-M. Duan conceived and designed the experiment. Y.-F. Pu, S. Zhang, N. Jiang, W. Chang and C. Li performed the experiment.  Y.-F. Pu and S. Zhang analyzed the data. Y.-K. Wu contributed materials/analysis tools.  Y.-F. Pu, S. Zhang, Y.-K. Wu and L.-M. Duan wrote the paper.

\section{Competing interests}

The authors declare that there are no competing interests.

\end{document}


\title{Supplementary Information: ``Experimental demonstration of memory-enhanced scaling for entanglement connection of quantum repeater segments''}

\author{Y.-F. Pu$^1$\footnote{%
These authors contributed equally to this work.\label{FN}}\footnote{Present address: Institute for Experimental Physics, University of Innsbruck, A-6020 Innsbruck, Austria}, S. Zhang$^{1}$\footnotemark[1], Y.-K. Wu$^1$, N. Jiang$^1$\footnote{Present address: Department of Physics, Beijing Normal University, Beijing 100875, China}, W. Chang$^1$, C. Li$^1$, L.-M. Duan}
\affiliation{Center for Quantum Information, IIIS, Tsinghua University, Beijing 100084,
PR China}

\maketitle

\subsection{ Section S1. Experimental setup and sequence}

Here we describe the experimental sequence in detail. As shown in Fig.~S1, we first load two Rubidium-87 atomic clouds in Magneto-optical traps (MOT) at both sides for $0.35$ s. The experimental setups at QN2 and QN3 are different, with a double-chamber MOT setup at QN2 and a single-chamber MOT at QN3, which leads to difference in the number of captured atoms and optical depth (OD) between the two setups. In order to load the atoms into 1-D optical lattice, we first compress the atomic clouds by ramping up the current in the MOT gradient coil for $20$ ms which reduces the size and increases the density of the atomic cloud, and then apply a molasses cooling of $20$ ms long (sub-Doppler cooling) to further cool down the temperature of atoms to $20\,\mu$K ($50\,\mu$K) at QN2 (QN3). About $5\times10^6$ ($4\times10^5$) atoms are loaded into a 1D optical lattice by interfering two circularly-polarized $1064\,$nm laser beams. The two lattice beams have Gaussian diameters of $230\,\mu$m (220$\,\mu$m), and intersect each other with an angle of $10^\circ$. The lifetime of atoms in the lattice is about $2$ seconds, and the resonant optical depth (OD) is about $50$ and $10$ for QN2 and QN3 respectively due to the different MOT configurations. After the lattice loading, a magnetic field along the $+$z axis is turned on, and an optical pumping beam propagating along the $-$x axis pumps the atoms into the upper clock state $|g\rangle$. After that, the cooling and re-pumping lasers and the magnetic field gradient are turned off and the untrapped atoms are allowed to freely fall out of the lattice region for $60$ ms. The $1064$ nm laser is permanently on during the whole experimental sequence. The elliptical optical pumping beam, consisting of two $\pi$-polarized transitions resonant with $|g\rangle \rightarrow |e\rangle$ and $|s\rangle \rightarrow |e\rangle$, pumps the atoms into the upper clock state $|g\rangle$ within $50 \mu$s. All the steps above are controlled by two National Instruments 6733 boards.

\begin{figure}
  \centering
  \includegraphics[width=14cm]{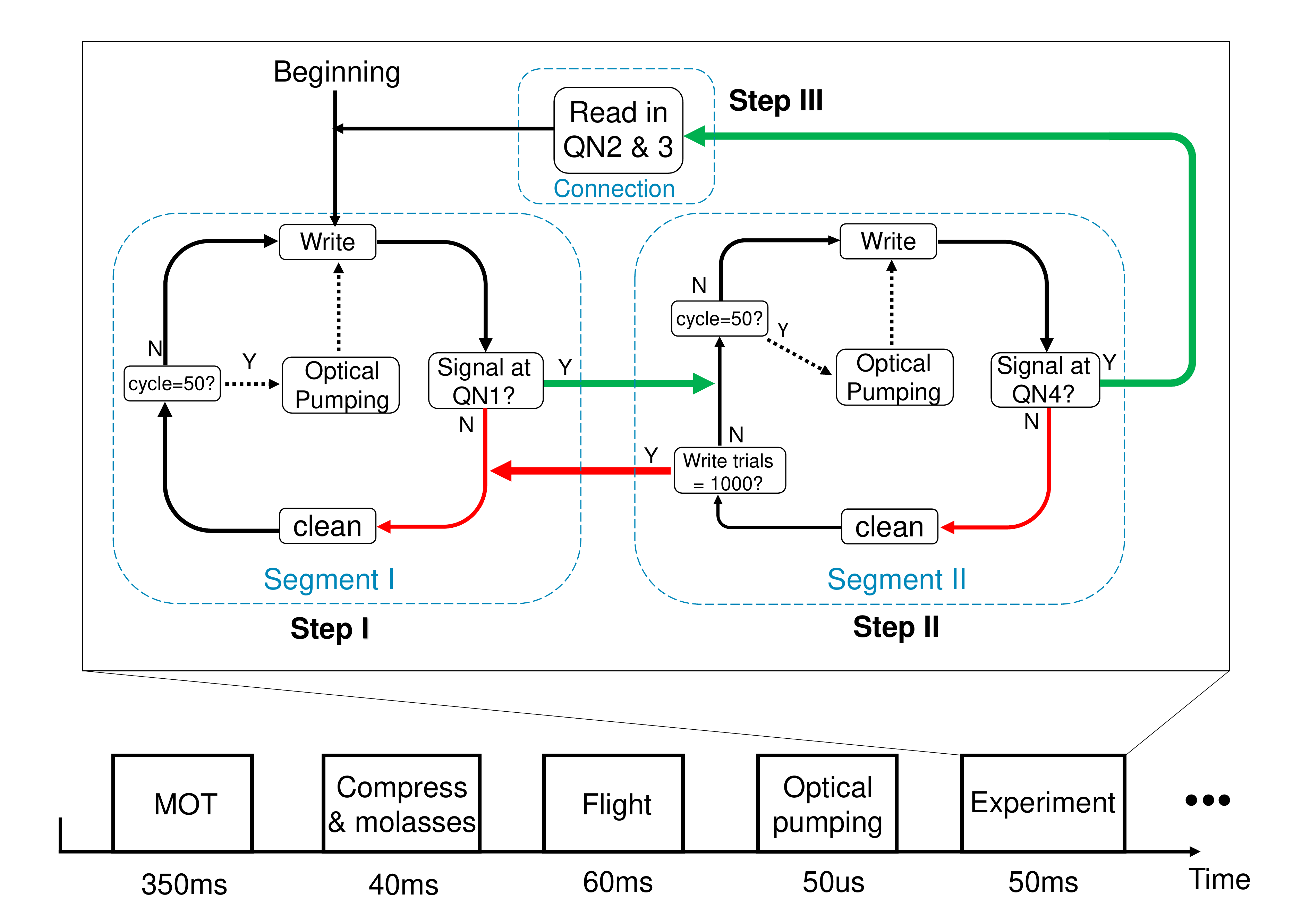}\\
  \caption{\textbf{The detailed experimental sequence.}
  }
\end{figure}

The detailed sequence of our protocol is illustrated in Fig.~S1. The time sequence in this part is controlled by a home-made-field-programmable gate array (FPGA), triggered by the 6733 board. The experimental sequence for entanglement generation and entanglement swapping begins with a write pulse of $100$ ns in segment I. If no signal photon is detected, we apply a clean pulse of 100 ns, which is identical to the read pulse, to pump the atoms back to initial state. It takes $1 \mu$s to complete the write-clean cycle for each trial. After every $50$ write-clean cycles, we optically pumped again for $5 \mu$s. The sequence enters write-clean cycle in segment II once a signal photon is registered at QN1. The cycle of segment II is basically the same as that of segment I, except that if there is no signal photon detected at QN4 after $1000$ trials, the sequence terminates and restarts the write-clean cycle of segment I. When both segments have successfully detected signal photons, we simultaneously apply read pulses on two memories to convert the spin-waves and direct two retrieved idler photons to the middle station for the joint Bell state measurement (BSM). After the BSM, the sequence restarts from the beginning.

To stabilize the phase difference of the two paths L and R, a $767\,$nm laser beam locked to a potassium vapor cell is inserted into the two paths with a dichroic mirror (DM) and is collected from the other end of the PBS2 for feedback, as shown in Fig.~2a. A piezo-driven prism in path L is employed for the active stabilization, but is not shown here for clarity. In addition, temperature-stabilized Fabry-Perot cavities are exploited for frequency filtering in each signal/idler path.


\subsection{ Section S2. EIT storage}

\begin{figure}
  \centering
  \includegraphics[width=12cm]{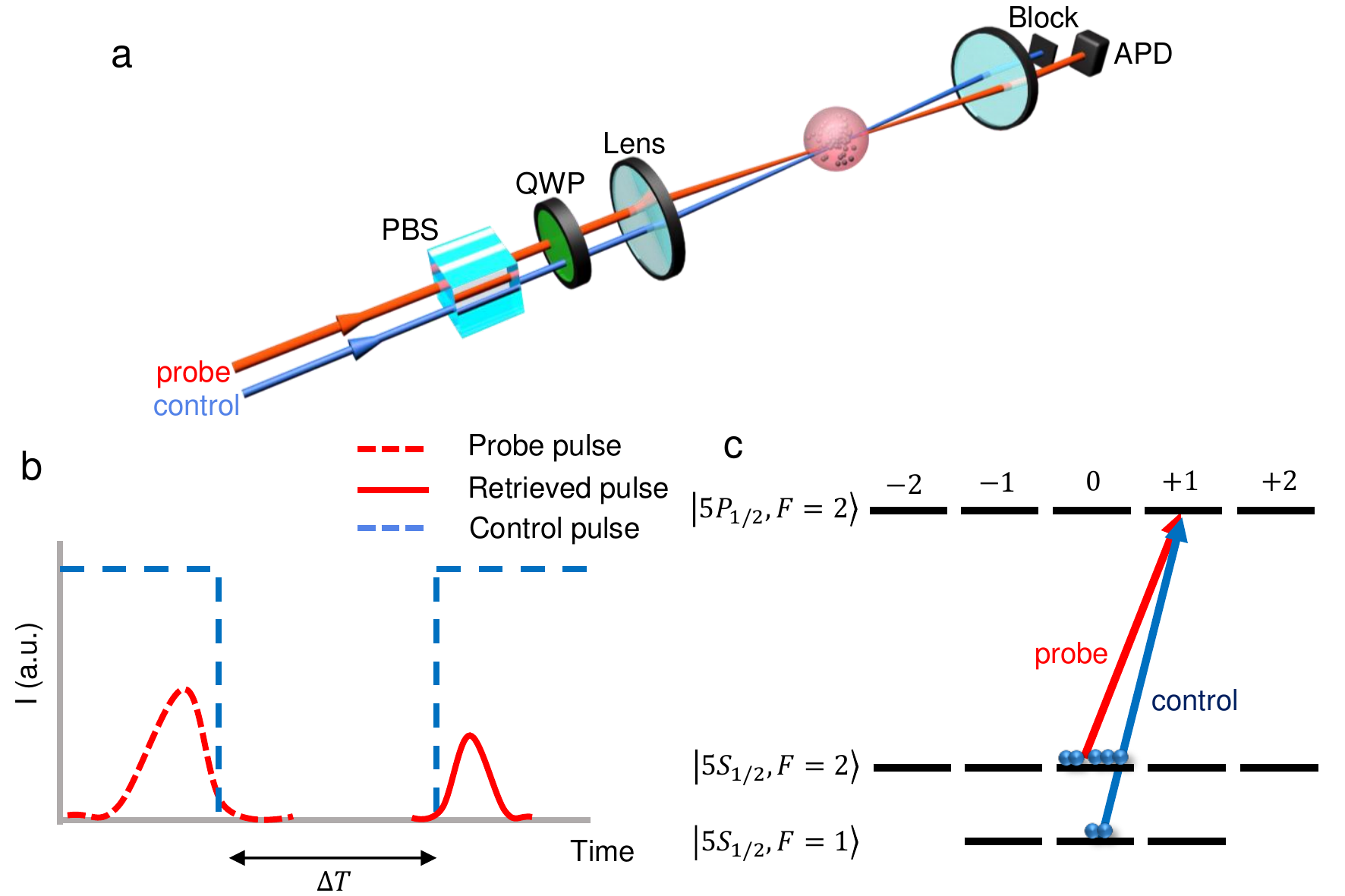}\\
  \caption{\textbf{The EIT storage setup.}
  \textbf{a,} The simplified setup of EIT storage. The optical lattice is on during the EIT storage but is not shown here for clarity. \textbf{b,} The time sequence for EIT storage. $I$ is the intensity on the detector. The red dashed pulse is the leaking probe pulse in the storage process, red solid pulse is the retrieved probe pulse after storage of time $\Delta T$, and the blue dashed line represents the control field intensity.  \textbf{c,} The energy level diagram for probe and control.}
\end{figure}

The simplified setup of EIT storage is illustrated in Fig.~S2a. A weak probe pulse resonant with the $|g\rangle\rightarrow|e\rangle\equiv|5P_{1/2},F=2,m_F=+1\rangle$ transition and the control field resonant with the $|s\rangle\rightarrow|e\rangle$ transition are used in the slow light configuration [27, 28], as shown in Fig.~S2c. The control beam propagates along the same direction of write/read and the probe beam is along one of the two spatial modes L or R, with an angle of $1.5^\circ$ to the control beam. Before the EIT storage, atoms are pumped into $|g\rangle$. The probe and the control beams have the same circular polarization and the probe pulse is stored in the $|g\rangle\leftrightarrow|s\rangle$ coherence when the control beam is turned off. After a variable time $\Delta T$, the control beam is turned on again, and the stored atomic coherence is converted back to the probe pulse which is measured on the detector, as illustrated in Fig.~S2b.

\subsection{ Section S3. The time-independent phase in the atom-photon entangled state}

\begin{figure}
  \centering
  \includegraphics[width=9cm]{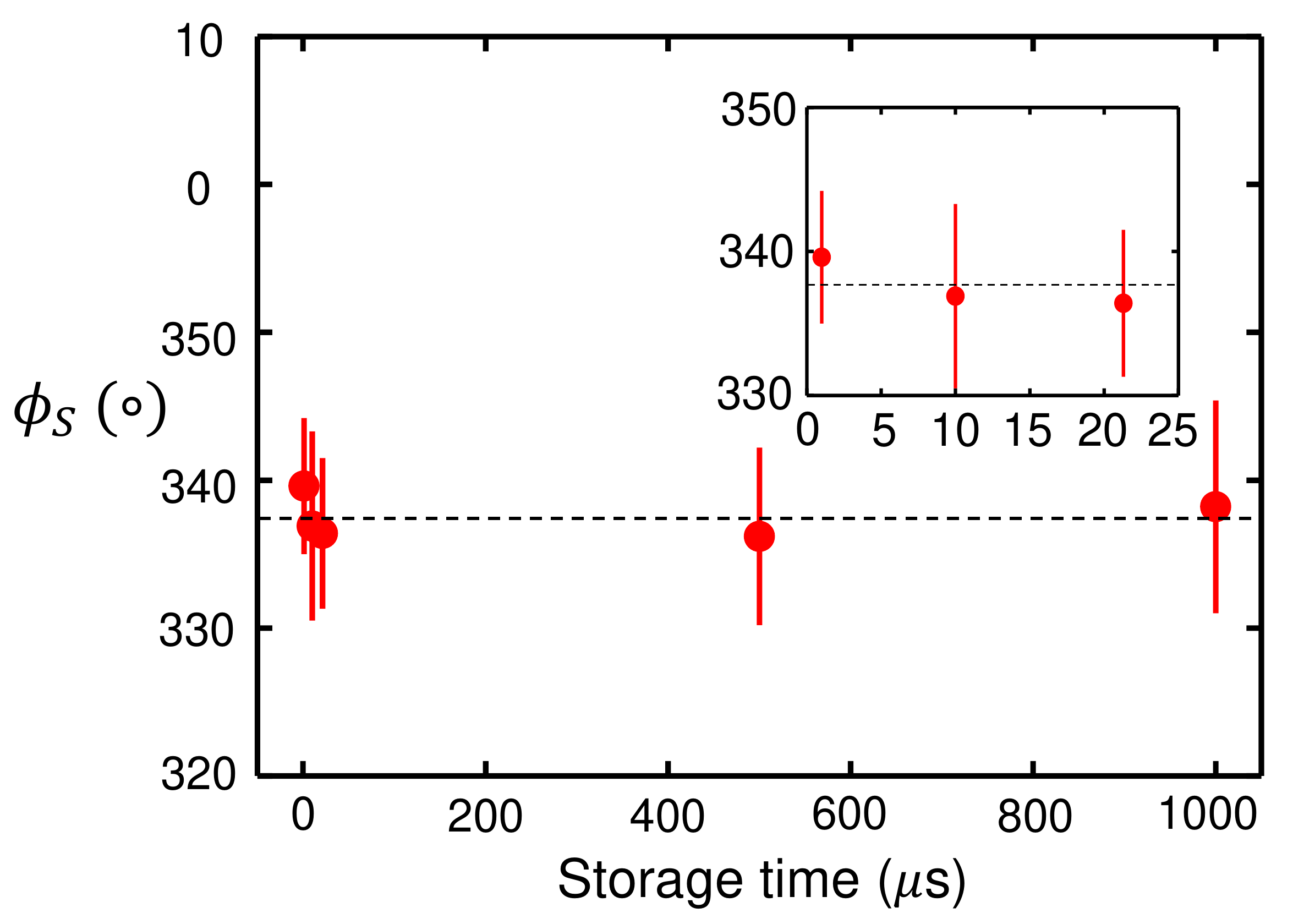}\\
  \caption{\textbf{The time-invariant phase in the atom-photon entangled state during storage.}
  The phase $\phi_S$ of the established atom-photon entanglement in segment I is measured at different storage times of $1 \mu$s, $10 \mu$s, $21.3 \mu$s, $500 \mu$s and $1000 \mu$s. The first $3$ phases are also shown in the inset. The black dashed line represents a constant phase at $\phi_S=337.5^\circ$. The error bars represent one standard deviation.}
\end{figure}

Here we demonstrate the phase $\phi_S$ (see Eq.(1) in the main text) in the atom-photon entangled state generated in step I is time-invariant during the storage in step II. We measure the phase $\phi_S$ at different storage times of $1 \mu$s, $10 \mu$s, $21.3 \mu$s, $500 \mu$s and $1000 \mu$s in segment I, and it remains constant up to a couple of degrees, as illustrated in Fig.~S3 and Tab. S1. This invariant phase is guaranteed by the two spatial bases $|L\rangle$ and $|R\rangle$ used here. The angles are extracted from the reconstructed density matrix from quantum tomography of the atom-photon entangled state in segment I. As a comparison, the phase of the atom-photon entangled state keeps increasing at a constant rate of $2\pi/1.2 \mu$s due to the Larmor precession in Ref. [30], where two Zeeman sub-levels are employed to be the stationary qubit bases.

\renewcommand{\arraystretch}{1.5}

\begin{table}[h!]
  \begin{center}
  \small
    \begin{tabularx}{1\textwidth} {
    >{\centering\arraybackslash}X
    >{\centering\arraybackslash}X
    >{\centering\arraybackslash}X
    >{\centering\arraybackslash}X
    >{\centering\arraybackslash}X
    >{\centering\arraybackslash}X }
    \hline
    \hline
    \textbf{Storage time ($\mu$s)} & $1 \mu$s & $10 \mu$s & $21.3 \mu$s & $500 \mu$s & $1000 \mu$s \\[1pt]
    \hline
    \textbf{$\phi_S$ $(^\circ)$} & 340(5) & 337(6) & 336(5) &  336(6) & 338(7)\\
    \hline
    \hline
    \end{tabularx}
  \end{center}
  \caption{$\phi_S$ at different storage times.}
\end{table}

\subsection{ Section S4. Expected four-photon coincidence rate}

In this section we calculate the expected rate of successful generation of entanglement between two signal photons. We assume the expected time to generate two signal photons in step I and II to be $T$, the time for each write trial is $A$, optical pumping time is $5A$ ($A=1\mu$s corresponds to our experimental parameters), the overall excitation probabilities on both nodes are $p$, and the maximum number of write operations at node B is $n$, then we have:
\begin{equation}
T=\underbrace{\sum_{i=1}^n(1-p)^{i-1}p(\overbrace{T'}^{\text{1st photon}} +iA+\overbrace{\lfloor \frac{i-1}{50} \rfloor 5A)}^{\text{optical pumping}}}_{\text{2nd photon is detected within n  trials}}+\underbrace{(1-p)^n(T'+nA+\lfloor\frac{n-1}{50}\rfloor5A+T)}_{\text{fail to detect 2nd photon, restart}}
\end{equation}

\noindent where $T'=\underset{i}{\sum}(1-p)^{i-1}p( iA+\lfloor \frac{i-1}{50} \rfloor 5A)$ is the expected time of generating $1$st photon. The overall excitation probability is $p=\chi\epsilon_s$, where $\chi$ is the intrinsic atom-photon entanglement excitation probability, $\epsilon_s\equiv\epsilon_s^f\epsilon_s^t\epsilon_s^d\approx0.33$ is the detection efficiency for a signal photon, which includes the fiber coupling efficiency $\epsilon_s^f=0.75$, optical transmission (including the etalon for frequency filtering) $\epsilon_s^t\approx0.8$, and detector efficiency $\epsilon_s^d\approx 0.55$. Here we introduce a correction factor $C$ which is the ratio between $T$ and the ideal expected preparation time $\frac{2A}{p}$ (where the trial number is not limited to $n$, and the optical pumping time is not considered.):
\begin{equation}
C(n,p)=T/\frac{2A}{p}
\end{equation}
In the case that $n=1000$ and $p> 0.1\%$, $C$ is within the range $(1, 1.42)$, as illustrated in Fig.~S4a.

The rate $R$ of generating four-photon coincidence is inversely related to the expected time $T$ which can be expressed as:
$R\propto\frac{1}{T}=\frac{p}{2A}\frac{1}{C}$.
 Taking into account the overall retrieval efficiency of idler photon $\eta_2\epsilon_i\approx4.2\%$, $\eta_3\epsilon_i\approx1.9\%$, (where $\eta_{2(3)}$ are the intrinsic retrieval efficiency of QN2(3), $\epsilon_i$ is the overall detection efficiency of idler photon. There's one more fiber in the optical path in the BSM than the setup of measuring DLCZ retrieval efficiency, thus another fiber coupling and transmission efficiency about 0.65 is put into $\epsilon_i$), A$=1\mu$s and the duty cycle $D\approx10\%$ (as illustrated in Fig.~S1), the four-fold coincidence rate can be expressed as:
\begin{equation}
R=\underbrace{\frac{\chi\epsilon_s}{2A}\frac{1}{C(p)}}_{\text{two signals}}\underbrace{\frac{\eta_2\eta_3\epsilon_i^2}{2}}_{\text{BSM}}D
\end{equation}


\noindent where $\chi\epsilon_s=p$ is the probability of detecting a signal photon in horizontal basis. Furthermore, we fit the curve $R$ to the measured data by slightly changing the idler photon efficiency. The fitted curve $R$ and the linear function which has the same expression of Eq.~S3 but with a fixed $C_{p=1\%}=1.08$ are illustrated in Fig.~S4b.

\begin{figure}
  \centering
  \includegraphics[width=18cm]{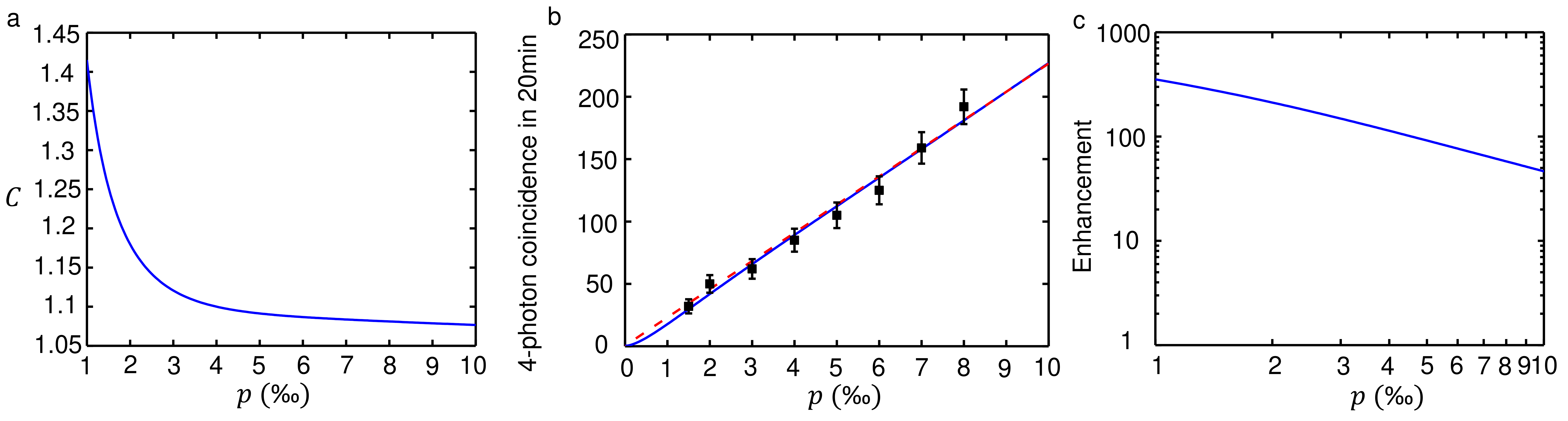}\\
  \caption{\textbf{The linear scaling in the efficiency of entanglement connection.}
  \textbf{a,} The correction coefficient $C$ vs $p$ , when $n=1000$. \textbf{b,} The four-photon coincidence counts in 20 minutes. The black squares are the measured data at different $p$, the blue curve is the Fitted $R$, and the red dashed line is a linear function modified from $R$ with the correction factor fixed to $C_{p=1\%}=1.08$. \textbf{c,} The memory enhancement $R/R'$ in the four-photon rate over the scheme without memory enhancement.}
\end{figure}

As a comparison, we also consider the four-photon coincidence rate without the memory enhancement. For the entanglement connection scheme without memory enhancement, synchronous generation of two atom-photon entanglement is required for the successful entanglement swapping. Synchronous generation of two atom-photon entanglement is heralded by simultaneously detecting two signal photons in both QN1 and QN4 in a single trial, and then entanglement swapping is performed on QN2 and QN3 to connect two quantum repeater segments. In this case, the 4-photon coincidence rate can be expressed as:
\begin{equation}
R'=\underbrace{\frac{(2\chi\epsilon_s)^2}{4A}}_{\text{heralding}} \underbrace{\frac{\eta_2\eta_3\epsilon_i^2}{2}}_{\text{idler}} D.
\end{equation}

As can be seen clearly, the four-photon coincidence generation rate $R$ in the memory-enhanced case is linearly dependent on the excitation probability $\chi$ (up to a correction factor $C<1.42$ within the range of $p>0.1\%$), and $R'$ is quadratically dependent on $\chi$ without the memory enhancement. With this we estimate the acceleration of our protocol to the two-photon heralding scheme by $R/R'=\frac{1}{2Cp}$, which is about $353$ times when $p=0.1\%$, as shown in Fig.~S4c.

\subsection{ Section S5. Estimated time cost for each step of the experimental protocol}

In this section we estimate the time cost in each step. With the excitation probability of both segments I and II set to $p=0.1\%$, the distribution of signal photon arrival time at QN1 (QN4) in step I (II) from the beginning of the corresponding step is illustrated in Fig.~S5a, with the arrival time of QN1 photon unlimited but the QN4 photon truncated at $N=1000$ by our protocol. Both the arrival time in two segments follow the geometric distribution where the signal photon is detected in $i$-th trial can be expressed as: $P(i)=(1-p)^{i-1}p$, as described in Eq.~(S1). In addition, we also simulate the distribution of the time cost of a full operation consisting of Step I, II and III (or the time separation of two successive step III), which is equivalent to the expected time cost of the entanglement preparation (the time cost of step III is $1\mu$s following step II immediately, negligible compared to step I and II), is demonstrated in Fig.~S5b with $p=0.1\%$. The average time cost of three steps and the ideal time cost $\frac{2}{p}$ are also marked in Fig.~S5b, and the ratio between them matches the correction factor $C$ described in Eq.~(S2). The corresponding distributions with $p=0.4\%$ are illustrated in Fig.~S5c and d.

We also simulate the average storage time QN2 experienced in every successful step II heralded by a photon click in QN4, which determines the slight fidelity change of the entanglement in segment I right before the entanglement connection. This average storage time can be directly retrieved from the truncated geometric distribution of the trial number in step II, as illustrated in Fig. S5a and c. Here we demonstrate the average storage time for the excitation probabilities varying from $p=0.1\%$ to $0.8\%$, as shown in Tab. S2.

Furthermore, we also demonstrate the average repetition number of step I and II before we arrive at step III (for example, this number is 3 if we fail and restart twice at step II) in Fig.~S6. It can be seen clearly that this repetition number decays to 1 very fast with increasing $p$.

\begin{figure}
  \centering
  \includegraphics[width=11.5cm]{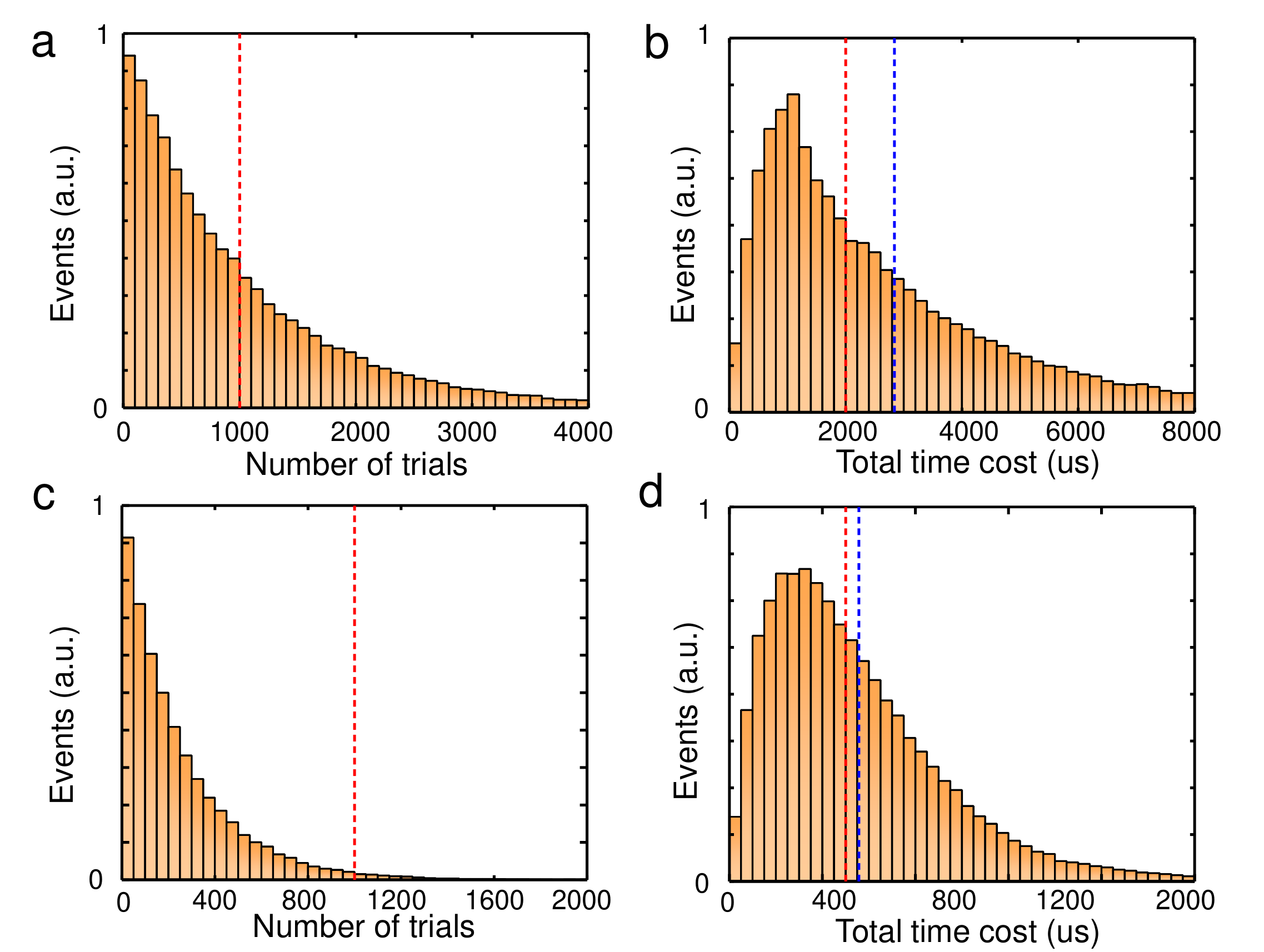}\\
  \caption{\textbf{The time cost of the experimental protocol.}
  \textbf{a,} The trial number distribution of step I/II with $p=0.1\%$. The red dashed line represents the maximally trial number N=1000 in step II. \textbf{b,} The total time cost distribution of step I, II and III with $p=0.1\%$. The red dashed line represents the ideal time cost $\frac{2}{p}$, and the blue dashed line is the average time cost. \textbf{c, d} The distribution of same variables in a and b, with $p=0.4\%$.}
\end{figure}
\begin{figure}
  \centering
  \includegraphics[width=7.8cm]{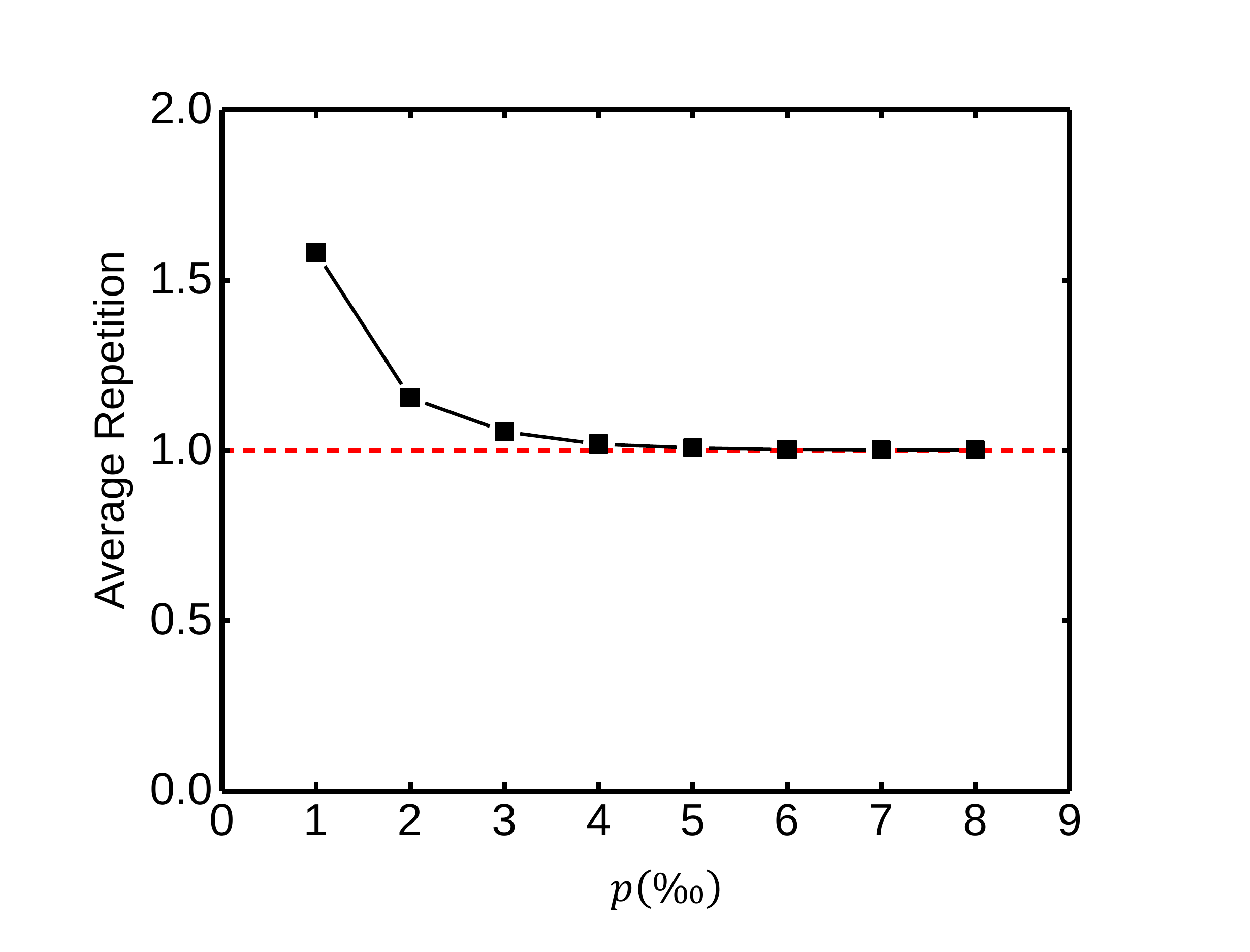}\\
  \caption{\textbf{The average repetition number of step I and II for a successful implementation of step III.}
  The average repetition number of step I and II to herald a step III is demonstrated with different $p$ varying from $0.1\%$ to $0.8\%$.}
\end{figure}

\begin{table}
\begin{center}
\begin{tabular}{ccccccccc}
  \hline
  \hline
  \textbf{Excitation probability}(\%)$\qquad$ & 0.1$\qquad$ & 0.2$\qquad$ & 0.3$\qquad$ & 0.4$\qquad$  & 0.5$\qquad$  & 0.6$\qquad$  & 0.7$\qquad$  & 0.8$\qquad$  \\ \hline
  \textbf{Average storage time}($\mu$s) $\qquad$ & 462 $\qquad$ & 377 $\qquad$ & 309$\qquad$ & 255$\qquad$ & 212$\qquad$ & 181$\qquad$ & 156$\qquad$ & 136$\qquad$ \\
  \hline
  \hline
\end{tabular}
\end{center}
\caption{\textbf{Average storage time in QN2 for a successful step II.}}
\end{table}

\subsection{Section S6. Discussion of the scaling-changing acceleration}

For an atomic-ensemble quantum memory, the excitation probability $p$ of the write pulse gives an intrinsic infidelity $p$ for the produced entangled state due to the double excitation. Therefore, the intrinsic quality of entanglement and the success probability are in a trade-off relation. Let us discuss this trade-off relation and its effect for the scaling change induced by the quantum repeaters. Let us take the two-segment case as an example where the transmission efficiency on each segment is $q$. For a fair comparison with direct communication, if one wants to fix the overall intrinsic infidelity to $p$ for two segments, the excitation probability for each segment should be $p/2$ and the corresponding success probability will be $pq/2$. For the memory-based quantum repeater approach, if one generates entanglement in the two segments sequentially and perform on-demand entanglement swapping when both segments succeed, the average total number of repetitions will be $4/(pq)$. For the direct communication over these two segments, the number of repetition will instead be $1/(p q^2)$. While if one performs simultaneous entanglement swapping without memory enhancement, the number of repetition will be $4/(p^2 q^2)$. Now one can see that with the same intrinsic infidelity $p$, the quantum repeaters improve the efficiency scaling from $O(1/q^2)$ to $O(1/q)$, while the requirement of a fixed total excitation probability only contributes a numerical factor of 2. It is this efficient scaling with respect to the transmission efficiency $q$ that is crucial for quantum repeaters in long-distance entanglement distribution, which will not be affected by tuning the excitation probability $p$. The mechanism for this improvement from a quadratic scaling to a linear one in the two-segment case is the on-demand entanglement swapping enabled by the quantum memory when both segments of entanglement have been heralded, and the same mechanism will lead to an improvement from an exponential scaling to a polynomial one for the multi-segment case under the same transmission efficiency with the direct communication.

The key idea behind the above scaling change is the on-demand entanglement swapping enabled by the quantum memory. Our experiment is targeted to demonstration of this on-demand entanglement swapping by use of the quantum memories. Note that for the above $4/(pq)$ scaling of memory-enhanced on-demand entanglement swapping and the $4/(p^2 q^2)$ scaling without memory enhancement, the excitation probability $p$ and the transmission efficiency $q$ play the same role. So instead of adjusting the transmission loss, we choose to tune the excitation probability $p$ in Fig.~4 of the main text to study the scaling. Furthermore, in our experiment, the intrinsic infidelity of the entanglement due to the excitation probability $p$ is about $p\sim 0.3\%$, which is negligible compared with the contribution of other technical noise to the infidelity of the final entangled state ($\sim 20\%$). So the trade-off relation between the intrinsic fidelity and the success probability does not influence the comparison here.

In Fig.~4 of the main text, what we compare is the efficiencies between entanglement swapping with quantum memory and entanglement swapping without the use of quantum memory. In our experiment, the infidelity induced by coherence decay in the quantum memory is only about 3\% for the required $1\,$ms maximal quantum storage time, which is much smaller than the overall infidelity ($\sim 20\%$) of the final sate after, dominated by the contribution of the technical noise in the entanglement swapping and the addition of the infidelities from the two-segment EPR states. However, by use of the quantum memory, the efficiency improves from a quadratic scaling with the segment entanglement preparation probability $p$ to a linear scaling. This is the key idea responsible for the efficiency improvement by memory-based quantum repeaters.